\documentclass{iopart}
\usepackage{epsfig}  

\def\lsim{\raise0.3ex\hbox{$<$\kern-0.75em\raise-1.1ex\hbox{$\sim$}}}
\def\gsim{\raise0.3ex\hbox{$>$\kern-0.75em\raise-1.1ex\hbox{$\sim$}}}

\begin{document}

\title{
DGLAP analyses of nPDF: constraints from data}

\author{  K.J.~Eskola\dag, H.~Honkanen\dag, V.J.~Kolhinen\dag, P.V.~Ruuskanen\dag, C.A. Salgado\ddag\footnote[3]{Invited talk presented by C.A. Salgado at the NuFact'02 workshop.},}
\address{\dag Department of Physics, University of Jyv\"askyl\"a,\\
P.O.Box 35, FIN-40014 University of Jyv\"askyl\"a (Finland)}
\address{\ddag CERN-TH, CH-1211 Geneva 23 (Switzerland).}

\begin{abstract}
We explain how the constraints from present experimental data can be used
to obtain 
the nPDF in the framework of LO DGLAP evolution. 
We will also  compare the only two available sets of this type and
comment on the important information that neutrino 
factories could provide.
\end{abstract}

\section{Introduction}
Parton distribution functions (PDF) are needed to compute
hard processes in hadronic and nuclear collisions. 
The  method to obtain the PDF from
experimental data is well established in the 
case of the free proton:
the initial
distributions at $Q_0^2$ are evolved by the DGLAP equations \cite{dglap}
to larger $Q^2$
and fitted to available data. 
The data from deep inelastic lepton-proton scattering (DIS) are
of main importance in these
analyses.
The nuclear structure functions $F_2^A$ measured in DIS experiments
differ from 
those of the free nucleons. Definining the ratio vs. deuterium,
$R_{F_2}^A(x,Q^2)= \frac{1}{A}F_2^A(x,Q^2)/(\frac{1}{2}F_2^D(x,Q^2)),$
several 
nuclear
effects can be distinguished: shadowing ($R_{F_2}^A <1$) at
small values of $x$, antishadowing ($R_{F_2}^A >1$)
at intermediate $x$ and EMC 
effect 
($R_{F_2}^A <1$ again)
and Fermi motion at large $x$.
The nuclear effects in $F_2^A$ translate in nuclear PDF (nPDF) which are modified from the ones of the free proton.
The goal then is to obtain a set of nPDF following
the well established procedure used for 
the free proton. 
In practice, a set of ratios of the PDF in bound and free protons, 
$R_i^A(x,Q^2)=f_i^A(x,Q^2)/f_i^p(x,Q^2)$ for
$i=g,\, u_V,\, d_V,\, \bar u\dots$, are extracted for a known set of the 
free proton PDF $f_i^p(x,Q^2)$.

\section{EKRS analysis} 

\begin{figure}
\begin{center}
\epsfxsize=8cm
\epsfbox{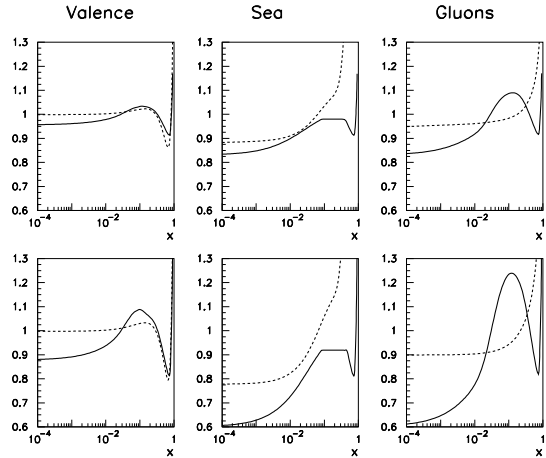}
\end{center}
\vspace{-0.5cm}
\caption{\label{figSalgado1}EKS98 (solid lines) and HKM (dashed lines) 
nuclear modifications for 
valence, sea and gluon 
distributions in Pb and C at $Q^2=2.25$~GeV$^2$.}
\end{figure}

The main problem in the nuclear case is the lack of experimental data. 
The DGLAP analysis of EKRS, which lead to the set EKS98 \cite{Eskola:1998iy},
uses several sets of DIS data on $F_2^A$ (see \cite{Eskola:1998iy} for the
refs.) and data on the Drell-Yan (DY) process measured in pA collisions 
\cite{Alde:im}.
Some other sets of data could be very helpful in constraining 
the nuclear effects for different parton flavours,
e.g. charm production to constrain gluons. However, so far they are
not included 
in \cite{Eskola:1998iy}
because of large 
error bars (open charm in $pA$) or the presence of final state nuclear 
effects (charmonium in $pA$).
Further constraints 
which are used
are momentum and baryon number sum rules.
At an initial scale, 
chosen as $Q_0^2=2.25$~GeV$^2$,
the ratios for valence quarks $R_V^A(x,Q^2_0)$
(same for $u_V^A$ and $d_V^A$), sea quarks 
$R_S^A(x,Q^2_0)$
(same for $\bar u^A$, $\bar d^A$ and $\bar s^A$)
and gluons $R_g^A(x,Q^2_0)$
are obtained
in the following way:

\begin{itemize}
\item At large values of $x$ ($x\gsim 0.3$) 
valence quarks dominate, and
the data on $R_{F_2}^A$ fix the ratio $R_V^A$
but do not constrain the ratio $R_S^A$.
There are no constraints for the nuclear gluons, either, in this
region. For consistency of the DGLAP evolution, it is assumed that
$R_S^A\approx R_V^A$, and that a similar EMC effect also exists in
$R_g^A$ already at $Q_0^2$.

\item At intermediate values of $x$ ($0.04\lsim x\lsim 0.3$) both DIS
and DY data constrain the ratios $R_V^A$ and $R_S^A$. The use of DY
data \cite{Alde:im} is essential in order to fix the relative 
strength of the valence and sea quark 
modifications,
as DIS alone cannot distinguish between them. 
The baryon number
sum rule imposes also constraints to 
$R_V^A$. 
The gluon ratio $R_g^A$ is constrained at $0.02\lsim x\lsim 0.2$ by
the NMC data on the $Q^2$ dependence of the ratio $F_2^{\rm Sn}/
F_2^{\rm C}$ \cite{Arneodo:1996ru}, and by momentum conservation.
A 20\% antishadowing is found for gluons at $x\sim 0.1$.

\item At small values of $x$ ($x\lsim 0.04$), $F_2$ is dominated by sea 
quarks, so DIS data 
constrains mainly $R_S^A$. The ratio
$R_V^A$ is fixed by 
baryon number conservation
and turns out to be larger  (less shadowing) than $R_S^A$. 
At $x\lsim 0.005$, where no information from data 
is obtained in the region $Q^2\gsim 1$~GeV$^2$, 
a saturation of the shadowing 
($R_{F_2}^A\rightarrow const.$) is assumed. This 
phenomenon
has been observed 
but only at $Q^2\ll1$~GeV$^2$. 
At the initial scale, $R_g^A\approx R_{F_2}^A$ is assumed
at $x\lsim0.01$, which leads to positive $\log Q^2$ slopes for 
$F_2^A$ (observed at $x\sim0.01$ \cite{Arneodo:1996ru}).

\end{itemize}
For a given initial condition, LO DGLAP  evolution is done. Then, comparing
with the data at different $Q^2$, the best initial distributions  are obtained 
through a recursive procedure. The resulting 
initial ratios at $Q_0^2$
can be seen in Figure 
\ref{figSalgado1}.

\section{Comparison with other approaches}

For the moment there is only one global DGLAP analysis on the nPDF
similar to EKRS \cite{Eskola:1998iy}, that of HKM  
\cite{Hirai:2001np,Kumano:2002ii}. 
Figure 1 shows a comparison of the EKS98 and HKM results for the
ratios $R_V^A$, $R_S^A$ and $R_g^A$ at $Q^2=2.25$~GeV$^2$.  The
difference between the results follows from the fact that the data
on the DY process \cite{Alde:im} and on the $Q^2$ dependence of $F_2^A$
\cite{Arneodo:1996ru} are not used as contraints in the HKM analysis.

The DY data set is
important in the EKRS analysis in fixing 
$R_V^A$ and $R_S^A$ at intermediate $x$: the DIS data forces $R_V^A > 1$
at $x\sim 0.1$ and, as the the DY cross sections show almost no nuclear 
effects (in $x_2$) there, the ratio $R_S^A$ is bound to be less than one
(no antishadowing for sea quarks).
In this conference, preliminary results from the HKM analysis 
with the DY data included were presented \cite{Kumano:2002ii}. 
As a result, a better agreement with EKS98 was found.

The $Q^2$-dependence of the structure function $F_2$ is very sensitive 
to the gluon distribution 
at small values of $x$. LO DGLAP evolution of the nPDF 
gives \cite{Eskola:2002us}
$\partial R_{F_2}^A(x,Q^2)/\partial 
\log Q^2 \sim R_g^A(2x,Q^2)-R_{F_2}^A(x,Q^2)$.
NMC has measured \cite{Arneodo:1996ru} positive 
$\log Q^2$
slopes 
for the ratio $F_2^{\rm Sn}/F_2^{\rm C}$. 
This implies that within the DGLAP framework
gluon shadowing cannot be 
much stronger 
than that in $F_2^A$
in the measured region $x\gsim 0.01$.
Also too weak a gluon shadowing is outruled \cite{Eskola:2002us}
by the NMC data.

\section{Improvements from neutrino DIS data} 

DIS experiments with neutrino and antineutrino projectiles could measure 
 \cite{Mangano:2001mj} 

\begin{eqnarray}
F_2^{\nu p}=2x\left(\bar u+d+s+\bar c\right)& \hskip 1cm
F_2^{\bar\nu p}=2x\left( u+\bar d+\bar s+c\right)\nonumber \\
xF_3^{\nu p}=2x\left(-\bar u+d+s-\bar c\right)& \hskip 1cm
xF_3^{\bar\nu p}=2x\left( u-\bar d-\bar s+c\right)\nonumber
\end{eqnarray}

\noindent with similar relations for neutrons.
Different flavors could then be disentangled
and  some of the uncertainties 
discussed
above (e.g. valence
at small $x$ and sea at large $x$) would 
become more directly
constrained by data. 
This would allow for a more detailed  analysis of the nPDF. 
Moreover, the valence/sea separation at medium $x$ would
be measured and could be compared with the results from
DY data. 
This would test the universality of the nPDF.
Measuring sea and valence quark distributions in $\nu A$ experiments
would also shed more light on some open questions of QCD in nuclei, 
such as the probability interpretation of the (n)PDF \cite{Brodsky:2002ue}.

\ack
CAS thanks S. Brodsky and M. Mangano for useful discussions.
CAS is supported by a Marie Curie Fellowship 
HPMF-CT-2000-01025.  Financial support from the
Academy of Finland, grant no. 50338, is gratefully acknowledged.

\end{document}